\documentclass[12pt]{article}
\usepackage{a4wide}
\usepackage{latexsym}
\usepackage{amsmath}
\usepackage{amsfonts}
\usepackage{amscd}
\usepackage{cite}
\usepackage{graphicx}
\usepackage{axodraw}

\usepackage{pslatex}
\usepackage[latin1]{inputenc}
\usepackage[T1]{fontenc}

\def\bq{\begin{eqnarray}}
\def\eq{\end{eqnarray}}

\def\eps{\varepsilon}

\begin{document}

\thispagestyle{empty}

\begin{flushright}
  MZ-TH/09-49 \\ 
  PITHA 09/35 
\end{flushright}

\vspace{1.5cm}

\begin{center}
  {\Large\bf Feynman graphs in perturbative quantum field theory\\
  }
  \vspace{1cm}
  {\large Christian Bogner$^1$ and Stefan Weinzierl$^2$\\
  \vspace{6mm}
      {\small \em $^1$ Institut f\"ur Theoretische Physik E, RWTH Aachen,}\\
      {\small \em   D - 52056 Aachen, Germany}\\
  \vspace{2mm}
      {\small \em $^2$ Institut f{\"u}r Physik, Universit{\"a}t Mainz,}\\
      {\small \em D - 55099 Mainz, Germany}\\
  } 
\end{center}

\vspace{2cm}

\begin{abstract}\noindent
  {
In this talk we discuss mathematical structures associated to Feynman graphs.
Feynman graphs are the backbone of calculations in perturbative quantum field theory. 
The mathematical structures -- apart from being of interest in their own right -- 
allow to derive algorithms for the computation of these graphs.
Topics covered are the relations of Feynman integrals to periods, shuffle algebras and 
multiple polylogarithms. 
   }
\end{abstract}

\vspace*{\fill}


\section{Introduction}

High-energy particle physics has become a field where precision measurements have become possible.
Of course, the increase in experimental precision has to be matched with more accurate calculations
from the theoretical side.
As theoretical calculations are done within perturbation theory, 
this implies the calculation of higher order corrections.
This in turn 
relies to a large extent on our abilities to compute Feynman loop integrals.
These loop calculations are complicated by the occurrence 
of ultraviolet and infrared singularities.
Ultraviolet divergences are related to the high-energy behaviour of the integrand.
Infrared divergences may occur if massless particles are present in the theory
and are related to the low-energy or collinear behaviour of the integrand.

Dimensional regularisation \cite{'tHooft:1972fi,Bollini:1972ui,Cicuta:1972jf}
is usually employed to regularise these singularities.
Within dimensional regularisation one considers the loop integral in $D$ space-time dimensions
instead of the usual four space-time dimensions.
The result is expanded as a Laurent series in the parameter $\eps=(4-D)/2$, describing the deviation
of the $D$-dimensional space from the usual four-dimensional space.
The singularities manifest themselves as poles in $1/\eps$.
Each loop can contribute a factor $1/\eps$ from the ultraviolet divergence and a factor $1/\eps^2$
from the infrared divergences. 
Therefore an integral corresponding to a graph with $l$ loops can have poles up to $1/\eps^{2l}$.

At the end of the day, all poles disappear: The poles related to ultraviolet divergences
are absorbed into renormalisation constants.
The poles related to infrared divergences cancel in the final result for infrared-safe observables, 
when summed over all degenerate states 
or are absorbed into universal parton distribution functions.
The sum over all degenerate states involves a sum over contributions with different
loop numbers and different numbers of external legs.

However, intermediate results are in general a Laurent series in $\eps$
and the task is to determine the coefficients of this Laurent series up to a certain order.
At this point mathematics enters. We can use the algebraic structures associated to Feynman integrals
to derive algorithms to calculate them.
A few examples where the use of algebraic tools has been essential are the calculation of the 
three-loop Altarelli-Parisi splitting functions 
\cite{Moch:2004pa,Vogt:2004mw}
or the calculation of the two-loop amplitude for the process $e^+ e^- \rightarrow \; \mbox{3 jets}$
\cite{Garland:2001tf,Garland:2002ak,Moch:2002hm,GehrmannDeRidder:2007bj,GehrmannDeRidder:2007hr,GehrmannDeRidder:2008ug,GehrmannDeRidder:2009dp,Weinzierl:2008iv,Weinzierl:2009ms,Weinzierl:2009yz}.

On the other hand is the mathematics encountered in these calculations of interest in its own right
and has led in the last years to a fruitful interplay between mathematicians and 
physicists.
Examples are the relation of Feynman integrals to mixed Hodge structures and motives, as well as
the occurrence of certain transcendental constants in the result of a calculation\cite{Bloch:2005,Bloch:2008jk,Bloch:2008,Brown:2008,Brown:2009a,Brown:2009b,Schnetz:2008mp,Schnetz:2009,Aluffi:2008sy,Aluffi:2008rw,Aluffi:2009b,Aluffi:2009a,Bergbauer:2009yu,Laporta:2002pg,Laporta:2004rb,Laporta:2008sx,Bailey:2008ib,Bierenbaum:2003ud}.

This article is organised as follows:
After a brief introduction into perturbation theory (sect.~\ref{sect:perturbation_theory}),
multi-loop integrals (sect.~\ref{sect:multi_loop}) and periods (sect.~\ref{sect:periods}),
we present in sect.~\ref{sect:theorem} a theorem stating that under rather weak assumptions the coefficients of the
Laurent series of any multi-loop integral are periods.
The proof is sketched in sect.~\ref{sect:sector_decomp} and sect.~\ref{sect:hironaka}.
Shuffle algebras are discussed in sect.~\ref{sect:shuffle}.
Sect.~\ref{sect:polylog} is devoted to multiple polylogarithms.
In sect.~\ref{sect:calc} we discuss how multiple polylogarithms emerge in the calculation of Feynman
integrals.
Finally, sect.~\ref{sect:conclusions} contains our conclusions.

\section{Perturbation theory}
\label{sect:perturbation_theory}

In high-energy physics experiments one is interested in scattering processes with two
incoming particles and $n$ outgoing particles.
Such a process is described by a scattering amplitude, which can be calculated in perturbation theory.
The amplitude has a perturbative expansion in the (small) coupling constant $g$:
\bq
\label{basic_perturbative_expansion}
 {\mathcal A}_n & = & g^n \left( {\mathcal A}_n^{(0)} + g^2 {\mathcal A}_n^{(1)} + g^4 {\mathcal A}_n^{(2)} + g^6 {\mathcal A}_n^{(3)} + ... \right).
\eq
To the coefficient ${\mathcal A}_n^{(l)}$ contribute Feynman graphs with $l$ loops and $(n+2)$ external legs.
The recipe for the computation of ${\mathcal A}_n^{(l)}$ is as follows: Draw first all Feynman diagrams with the given number
of external particles and $l$ loops. Then translate each graph into a mathematical formula with the help of the Feynman
rules.
${\mathcal A}_n^{(l)}$ is then given as the sum of all these terms.

Feynman rules allow us to translate a Feynman graph into a mathematical formula.
These rules are derived from the fundamental Lagrange density of the theory, 
but for our purposes it is sufficient to accept them as a starting point.
The most important ingredients are internal propagators, vertices and external lines.
For example, the rules for the propagators of a fermion or a massless gauge boson read
\bq
\mbox{Fermion:}\;\;\;
\begin{picture}(50,20)(0,10)
 \ArrowLine(50,15)(0,15)
\end{picture} & = &
 i \frac{p\!\!\!/ +m }{p^2-m^2+i\delta},
 \nonumber \\
\mbox{Gauge boson:}\;\;\;
\begin{picture}(50,20)(0,10)
 \Photon(0,15)(50,15){4}{4}
\end{picture} & = &
 \frac{-i g_{\mu \nu}}{k^2+i\delta}.
\eq
Here $p$ and $k$ are the momenta of the fermion and the boson, respectively. $m$ is the mass of the fermion.
$p\!\!\!/=p_\mu \gamma^\mu$ is a short-hand notation for the contraction of the momentum with the Dirac matrices.
The metric tensor is denoted by $g_{\mu\nu}$ and the convention adopted here is to take the metric tensor as
$g_{\mu\nu} = \mbox{diag}(1,-1,-1,-1)$.
The propagator would have a pole for $p^2=m^2$, or phrased differently for $E=\pm \sqrt{\vec{p}^2+m^2}$.
When integrating over $E$, the integration contour has to be deformed to avoid these two poles.
Causality dictates into which directions the contour has to be deformed. 
The pole on the negative real axis is avoided
by escaping into the lower complex half-plane, the pole at the positive real axis is avoided by a deformation 
into the upper complex half-plane. Feynman invented the trick to add a small imaginary part $i\delta$ to the 
denominator, which keeps track of the directions into which the contour has to be deformed.
In the following we will usually suppress the $i\delta$-term in order to keep the notation compact.

As a typical example for an interaction vertex let us look at the vertex involving a fermion pair and a gauge boson:
\bq
\begin{picture}(50,30)(0,15)
 \Photon(0,20)(30,20){4}{2}
 \Vertex(30,20){2}
 \ArrowLine(30,20)(50,40)
 \ArrowLine(50,0)(30,20)
\end{picture} & = &
 i g \gamma^{\mu}.
 \\
 & & \nonumber 
\eq
Here, $g$ is the coupling constant and $\gamma^\mu$ denotes the Dirac matrices.
At each vertex, we have momentum conservation: The sum of the incoming momenta equals the sum of the outgoing momenta.

To each external line we have to associate a factor, which describes the polarisation of the corresponding particle:
There is a polarisation vector $\eps^\mu(k)$ for each external gauge boson and a spinor 
$\bar{u}(p)$, $u(p)$, $v(p)$ or $\bar{v}(p)$ for each external fermion.

Furthermore there are a few additional rules: First of all, there is an
integration
\bq
 \int \frac{d^4k}{(2\pi)^4}
\eq
for each loop. Secondly, each closed fermion loop gets an extra factor of $(-1)$.
Finally, each diagram gets multiplied by a symmetry factor $1/S$,
where $S$ is the order of the permutation group
of the internal lines and vertices leaving the diagram unchanged when the external lines are fixed.

Having stated the Feynman rules, let us look at two examples:
The first example is a scalar two-point one-loop integral with zero 
external momentum: 
\bq
\begin{picture}(100,40)(0,30)
 \Line(10,35)(25,35)
 \Line(75,35)(90,35)
 \CArc(50,35)(25,0,360)
 \Vertex(25,35){2}
 \Vertex(75,35){2}
 \Text(5,35)[r]{\scriptsize $p=0$}
 \Text(50,55)[t]{\scriptsize $k$}
 \Text(50,5)[t]{\scriptsize $k$}
\end{picture}
 & = &
\int \frac{d^4k}{(2\pi)^4} \frac{1}{(k^2)^2} 
 =
\frac{1}{(4\pi)^2} \int\limits_0^\infty dk^2 \frac{1}{k^2} = 
\frac{1}{(4\pi)^2} \int\limits_0^\infty \frac{dx}{x}.
\\
\nonumber
\eq
This integral diverges at $k^2\rightarrow \infty$ as well as at $k^2\rightarrow 0$.
The former divergence is called ultraviolet divergence, the later is called infrared divergence.
Any quantity, which is given by a divergent integral, is of course an ill-defined quantity.
Therefore the first step is to make these integrals well-defined by introducing a regulator.
There are several possibilities how this can be done, but the
method of dimensional regularisation 
\cite{'tHooft:1972fi,Bollini:1972ui,Cicuta:1972jf}
has almost become a standard, as the calculations in this regularisation
scheme turn out to be the simplest.
Within dimensional regularisation one replaces the four-dimensional integral over the loop momentum by an
$D$-dimensional integral, where $D$ is now an additional parameter, which can be a non-integer or
even a complex number.
We consider the result of the integration as a function of $D$ and we are interested in the behaviour of this 
function as $D$ approaches $4$.
The original divergences will then show up as poles in the Laurent series in $\eps=(4-D)/2$.

\begin{figure}
\includegraphics[bb= 80 635 535 725]{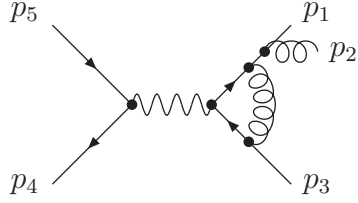}
\caption{\label{fig_ee_qqg} A one-loop Feynman diagram contributing to the process
$e^+ e^- \rightarrow q g \bar{q}$.}
\end{figure}

As a second example we consider
a Feynman diagram contributing to the one-loop corrections
for the process $e^+ e^- \rightarrow q g \bar{q}$, shown in fig.~\ref{fig_ee_qqg}.
At high energies we can ignore the masses of the electron and the light quarks.
From the Feynman rules one obtains for this diagram (ignoring coupling and colour prefactors):
\bq
\label{feynmanrules}
- \bar{v}(p_4) \gamma^\mu u(p_5)
  \frac{1}{p_{123}^2}
  \int \frac{d^{D}k_1}{(2\pi)^{4}}
  \frac{1}{k_2^2}
  \bar{u}(p_1) \eps\!\!\!/(p_2) \frac{p\!\!\!/_{12}}{p_{12}^2}
  \gamma_\nu \frac{k\!\!\!/_1}{k_1^2}
  \gamma_\mu \frac{k\!\!\!/_3}{k_3^2}
  \gamma^\nu
  v(p_3).
\eq
Here, $p_{12}=p_1+p_2$, $p_{123}=p_1+p_2+p_3$, $k_2=k_1-p_{12}$, $k_3=k_2-p_3$.
Further $\eps\!\!\!/(p_2) = \gamma_\tau \eps^\tau(p_2)$, where $\eps^\tau(p_2)$ is the
polarisation vector of the outgoing gluon.
All external momenta are assumed to be
massless: $p_i^2=0$ for $i=1..5$.
We can reorganise this formula into a part, which depends on the loop integration and a part, which does not.
The loop integral to be calculated reads:
\bq
\label{loop_int_example_1}
  \int \frac{d^D k_1}{(2\pi)^{4}}
  \frac{k_1^\rho k_3^\sigma}{k_1^2 k_2^2 k_3^2},
\eq
while the remainder, which is independent of the loop integration is given by
\bq
\label{loop_int_example_remainder}
- \bar{v}(p_4) \gamma^\mu u(p_5)
  \frac{1}{p_{123}^2 p_{12}^2}
  \bar{u}(p_1) \eps\!\!\!/(p_2) p\!\!\!/_{12}
  \gamma_\nu \gamma_\rho
  \gamma_\mu \gamma_\sigma
  \gamma^\nu
  v(p_3).
\eq
The loop integral in eq.~(\ref{loop_int_example_1}) contains in the denominator three propagator factors
and in the numerator two factors of the loop momentum.
We call a loop integral, in which the loop momentum occurs also in the numerator a ``tensor integral''.
A loop integral, in which the numerator is independent of the loop momentum is called a ``scalar integral''.
The scalar integral associated to eq.~(\ref{loop_int_example_1}) reads
\bq
\label{loop_int_example_1a}
  \int \frac{d^D k_1}{(2\pi)^{4}}
  \frac{1}{k_1^2 k_2^2 k_3^2}.
\eq
There is a general method \cite{Tarasov:1996br,Tarasov:1997kx}
which allows to reduce any tensor integral to a combination of scalar
integrals at the expense of introducing higher powers of the propagators and shifted space-time
dimensions.
Therefore it is sufficient to focus on scalar integrals.
Each integral can be specified by its topology, its value for the dimension $D$ and 
a set of indices, denoting the powers of the propagators.

\section{Multi-loop integrals}
\label{sect:multi_loop}

Let us now consider a generic scalar $l$-loop integral $I_G$ 
in $D=2m-2\eps$ dimensions with $n$ propagators,
corresponding to a graph $G$.
For each internal line $j$ the corresponding propagator
in the integrand can be raised to a power $\nu_j$.
Therefore the integral will depend also on the numbers $\nu_1$,...,$\nu_n$.
It is sufficient to consider only the case, where all exponents are natural numbers: $\nu_j \in {\mathbb N}$.
We define the Feynman integral by
\bq
\label{eq0}
I_G  & = &
 \frac{\prod\limits_{j=1}^{n}\Gamma(\nu_j)}{\Gamma(\nu-lD/2)}
 \left( \mu^2 \right)^{\nu-l D/2}
 \int \prod\limits_{r=1}^{l} \frac{d^Dk_r}{i\pi^{\frac{D}{2}}}\;
 \prod\limits_{j=1}^{n} \frac{1}{(-q_j^2+m_j^2)^{\nu_j}},
\eq
with $\nu=\nu_1+...+\nu_n$.
$\mu$ is an arbitrary scale, called the renormalisation scale.
The momenta $q_j$ of 
the propagators are linear combinations of the external momenta and the loop
momenta.
The prefactors are chosen such that after Feynman parametrisation the Feynman integral
has a simple form:
\bq
\label{eq1}
I_G  & = &
 \left( \mu^2 \right)^{\nu-l D/2}
 \int\limits_{x_j \ge 0}  d^nx \;
 \delta(1-\sum_{i=1}^n x_i)\,
 \left( \prod\limits_{j=1}^n x_j^{\nu_j-1} \right)
 \frac{{\mathcal U}^{\nu-(l+1) D/2}}{{\mathcal F}^{\nu-l D/2}}.
\eq
The functions ${\mathcal U}$ and $\mathcal F$ depend on the Feynman parameters
and can be derived
from the topology of the corresponding Feynman graph $G$.
Cutting $l$ lines of a given connected $l$-loop graph such that it becomes a connected
tree graph $T$ defines a chord ${\mathcal C}(T,G)$ as being the set of lines 
not belonging to this tree. The Feynman parameters associated with each chord 
define a monomial of degree $l$. The set of all such trees (or 1-trees) 
is denoted by ${\mathcal T}_1$.  The 1-trees $T \in {\mathcal T}_1$ define 
${\mathcal U}$ as being the sum over all monomials corresponding 
to the chords ${\mathcal C}(T,G)$.
Cutting one more line of a 1-tree leads to two disconnected trees $(T_1,T_2)$, or a 2-tree.
${\mathcal T}_2$ is the set of all such  pairs.
The corresponding chords define  monomials of degree $l+1$. Each 2-tree of a graph
corresponds to a cut defined by cutting the lines which connected the two now disconnected trees
in the original graph. 
The square of the sum of momenta through the cut lines 
of one of the two disconnected trees $T_1$ or $T_2$
defines a Lorentz invariant
\bq
s_{T} & = & \left( \sum\limits_{j\in {\mathcal C}(T,G)} p_j \right)^2.
\eq   
The function ${\mathcal F}_0$ is the sum over all such monomials times 
minus the corresponding invariant. The function ${\mathcal F}$ is then given by ${\mathcal F}_0$ plus an additional piece
involving the internal masses $m_j$.
In summary, the functions ${\mathcal U}$ and ${\mathcal F}$ are obtained from the graph as follows:
\bq
\label{eq0def}	
 {\mathcal U} 
 & = & 
 \sum\limits_{T\in {\mathcal T}_1} \Bigl[\prod\limits_{j\in {\mathcal C}(T,G)}x_j\Bigr]\;,
 \nonumber\\
 {\mathcal F}_0 
 & = & 
 \sum\limits_{(T_1,T_2)\in {\mathcal T}_2}\;\Bigl[ \prod\limits_{j\in {\mathcal C}(T_1,G)} x_j \Bigr]\, (-s_{T_1})\;,
 \nonumber\\
 {\mathcal F} 
 & = &  
 {\mathcal F}_0 + {\mathcal U} \sum\limits_{j=1}^{n} x_j m_j^2\;.
\eq

\section{Periods}
\label{sect:periods}

Periods are special numbers.
Before we give the definition, let us start with some sets of numbers:
The natural numbers $\mathbb{N}$,
the integer numbers $\mathbb{Z}$,
the rational numbers $\mathbb{Q}$,
the real numbers $\mathbb{R}$ and 
the complex numbers $\mathbb{C}$
are all well-known. More refined is already the set of algebraic numbers, 
denoted by $\bar{\mathbb{Q}}$.
An algebraic number is a solution of a polynomial equation with rational
coefficients:
\bq
 x^n + a_{n-1} x^{n-1} + \cdots + a_0 = 0,
 \;\;\; a_j \in \mathbb{Q}.
\eq
As all such solutions lie in $\mathbb{C}$, the set of algebraic numbers $\bar{\mathbb{Q}}$ 
is a sub-set of
the complex numbers $\mathbb{C}$.
Numbers which are not algebraic are called transcendental.
The sets $\mathbb{N}$, $\mathbb{Z}$, $\mathbb{Q}$ and $\bar{\mathbb{Q}}$ are countable, whereas
the sets $\mathbb{R}$, $\mathbb{C}$ and the set of transcendental numbers are uncountable.

Periods are a countable set of numbers, lying between $\bar{\mathbb{Q}}$ and $\mathbb{C}$.
There are several equivalent definitions for periods.
Kontsevich and Zagier gave the following definition \cite{Kontsevich:2001}:
A period is a complex number whose real and imaginary parts are values
of absolutely convergent integrals of rational functions with rational coefficients,
over domains in $\mathbb{R}^n$ given by polynomial inequalities with rational coefficients.
Domains defined by polynomial inequalities with rational coefficients
are called semi-algebraic sets.

We denote the set of periods by $\mathbb{P}$. The algebraic numbers are contained in the set of periods:
$\bar{\mathbb{Q}} \in \mathbb{P}$.
In addition, $\mathbb{P}$ contains transcendental numbers, an example for such a number is $\pi$:
\bq
 \pi & = & \iint\limits_{x^2+y^2\le1} dx \; dy.
\eq
The integral on the r.h.s. clearly shows that $\pi$ is a period.
On the other hand, it is conjectured that the basis of the natural logarithm $e$
and Euler's constant $\gamma_E$
are not periods.
Although there are uncountably many numbers, which are not periods, only very recently an example
for a number which is not a period has been found \cite{Yoshinaga:2008}.

We need a few basic properties of periods:
The set of periods $\mathbb{P}$ is a $\bar{\mathbb{Q}}$-algebra \cite{Kontsevich:2001,Friedrich:2005}.
In particular the sum and the product of two periods are again periods.

The defining integrals of periods have integrands, which are rational
functions with rational coefficients.
For our purposes this is too restrictive, as we will encounter
logarithms as integrands as well.
However any logarithm of a rational function with rational coefficients can be written as
\bq
 \ln g(x) 
 & = &
 \int\limits_0^1 dt \; \frac{g(x)-1}{(g(x)-1) t + 1}.
\eq

\section{A theorem on Feynman integrals}
\label{sect:theorem}

Let us consider a general scalar multi-loop integral as in eq.~(\ref{eq1}).
Let $m$ be an integer and set $D=2 m - 2 \eps$. Then this integral has 
a Laurent series expansion in $\eps$
\bq
 I_G & = & \sum\limits_{j=-2l}^\infty c_j \eps^j.
\eq
{\bf Theorem 1}: In the case where
\begin{enumerate}
\item all kinematical invariants $s_T$ are zero or negative, 
\item all masses $m_i$ and $\mu$ are zero or positive ($\mu\neq0$),
\item all ratios of invariants and masses are rational,
\end{enumerate}
the coefficients $c_j$ of the Laurent expansion are periods.

In the special case were
\begin{enumerate}
\item the graph has no external lines or all invariants $s_T$ are zero,
\item all internal masses $m_j$ are equal to $\mu$,
\item all propagators occur with power $1$, i.e. $\nu_j=1$ for all $j$,
\end{enumerate}
the Feynman parameter integral reduces to
\bq
I_G  & = &
 \int\limits_{x_j \ge 0}  d^nx \;
 \delta(1-\sum_{i=1}^n x_i)\,
 {\mathcal U}^{- D/2}
\eq
and only the polynomial ${\cal U}$ occurs in the integrand.
In this case it has been shown by Belkale and Brosnan \cite{Belkale:2003} that the coefficients of the 
Laurent expansion are periods.

Using the method of sector decomposition we are able to prove the general case \cite{Bogner:2007mn}.
We will actually prove a stronger version of theorem 1.
Consider the following integral
\bq
\label{basic_integral2}
J & = &
 \int\limits_{x_j \ge 0} d^nx \;\delta(1-\sum_{i=1}^n x_i)
 \left( \prod\limits_{i=1}^n x_i^{a_i+\eps b_i} \right)
 \prod\limits_{j=1}^r \left[ P_j(x) \right]^{d_j+\eps f_j}.
\eq
The integration is over the standard simplex.
The $a$'s, $b$'s, $d$'s and $f$'s are integers.
The $P$'s are polynomials in the variables $x_1$, ..., $x_n$ with rational coefficients.
The polynomials are required to be non-zero
inside the integration region, but
may vanish on the boundaries of the integration region.
To fix the sign, let us agree that all polynomials are positive inside the integration region.
The integral $J$ has a Laurent expansion
\bq
 J & = & \sum\limits_{j=j_0}^\infty c_j \eps^j.
\eq
{\bf Theorem 2}: The coefficients $c_j$ of the Laurent expansion of the integral $J$ are periods.

Theorem 1 follows then from theorem 2 as the special case
$a_i=\nu_i-1$, $b_i=0$, $r=2$, $P_1={\cal U}$, $P_2={\cal F}$, 
$d_1+\eps f_1 = \nu-(l+1)D/2$ and $d_2+\eps f_2 = l D/2 - \nu$.

Proof of theorem 2:
To prove the theorem we will give an algorithm which expresses each coefficient $c_j$
as a sum of absolutely convergent integrals over the unit hypercube with integrands,
which are linear combinations 
of products of rational functions with logarithms of rational functions,
all of them with rational coefficients.
Let us denote this set of functions to which the integrands belong by ${\cal M}$.
The unit hypercube is clearly a semi-algebraic set.
It is clear that absolutely convergent integrals
over semi-algebraic sets with integrands from the set ${\cal M}$ are periods.
In addition, the sum of periods is again a period.
Therefore it is sufficient to express each coefficient $c_j$ as a finite sum
of absolutely convergent integrals over the unit hypercube with integrands from ${\cal M}$.
To do so, we use iterated sector decomposition. 
This is a constructive method. Therefore we obtain as a side-effect a general purpose algorithm
for the numerical evaluation of multi-loop integrals.

\section{Sector decomposition}
\label{sect:sector_decomp}

In this section we review the algorithm for iterated
sector decomposition \cite{Hepp:1966eg,Roth:1996pd,Binoth:2000ps,Binoth:2003ak,Bogner:2007cr,Smirnov:2008py,Smirnov:2008aw}.
The starting point is an integral of the form
\bq
\label{basic_integral}
 \int\limits_{x_j \ge 0} d^nx \;\delta(1-\sum_{i=1}^n x_i)
 \left( \prod\limits_{i=1}^n x_i^{\mu_i} \right)
 \prod\limits_{j=1}^r \left[ P_j(x) \right]^{\lambda_j},
\eq
where $\mu_i=a_i+\eps b_i$ and $\lambda_j=c_j+\eps d_j$.
The integration is over the standard simplex.
The $a$'s, $b$'s, $c$'s and $d$'s are integers.
The $P$'s are polynomials in the variables $x_1$, ..., $x_n$.
The polynomials are required to be non-zero
inside the integration region, but
may vanish on the boundaries of the integration region.
The algorithm consists of the following steps:
\\
\\
Step 0: Convert all polynomials to homogeneous polynomials.
\\
\\
Step 1: Decompose the integral into $n$ primary sectors.
\\
\\
Step 2: Decompose the sectors iteratively into sub-sectors until each of the polynomials 
is of the form
\bq
\label{monomialised}
 P & = & x_1^{m_1} ... x_n^{m_n} \left( c + P'(x) \right),
\eq
where $c\neq 0$ and $P'(x)$ is a polynomial in the variables $x_j$ without a constant term.
In this case the monomial prefactor $x_1^{m_1} ... x_n^{m_n}$ can be factored out
and the remainder contains a non-zero constant term.
To convert $P$ into the form~(\ref{monomialised}) one chooses a subset
$S=\left\{ \alpha_{1},\,...,\, \alpha_{k}\right\} \subseteq \left\{ 1, \,...\, n \right\}$
according to a strategy discussed in the next section.
One decomposes the $k$-dimensional hypercube into $k$ sub-sectors according to
\bq
\label{decomposition}
 \int\limits_{0}^{1} d^{n}x & = & 
 \sum\limits_{l=1}^{k} 
 \int\limits_{0}^{1} d^{n}x
   \prod\limits_{i=1, i\neq l}^{k}
   \theta\left(x_{\alpha_{l}}\geq x_{\alpha_{i}}\right).
\eq
In the $l$-th sub-sector one makes for each element of $S$ the
substitution
\bq
\label{substitution}
x_{\alpha_{i}} & = & x_{\alpha_{l}} x_{\alpha_{i}}' \;\;\;\mbox{for}\; i\neq l.
\eq
This procedure is iterated, until all polynomials are of the form~(\ref{monomialised}).
\begin{figure}[t]
\includegraphics[bb= 100 440 570 710,width=0.9\textwidth]{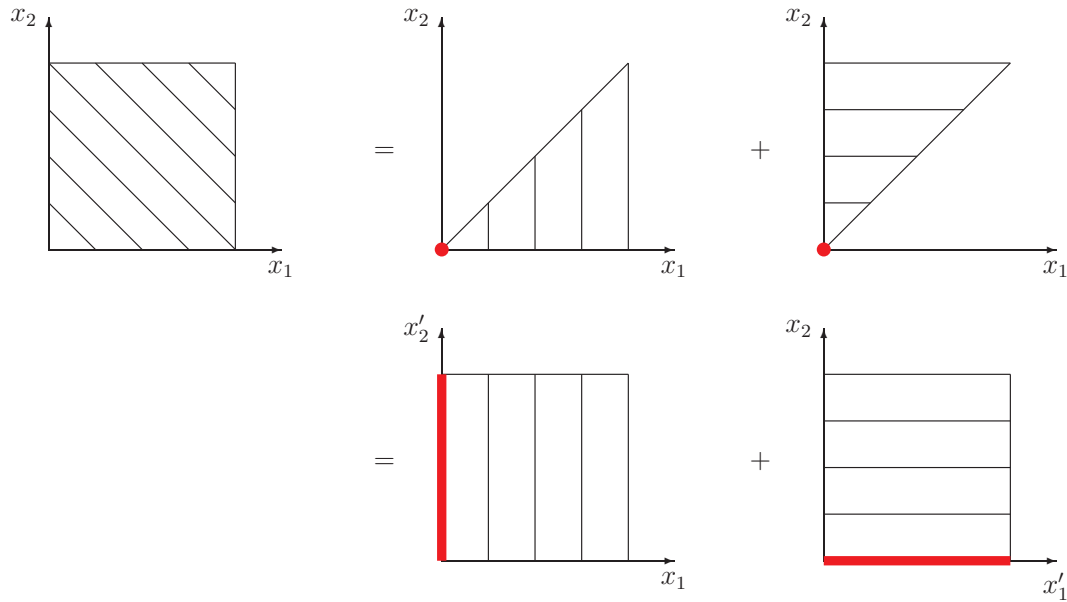}
\caption{\label{fig1} Illustration of sector decomposition and blow-up for a simple example.}
\end{figure}
Fig.~\ref{fig1} illustrates this for the simple example $S=\{1,2\}$. 
Eq.~(\ref{decomposition}) gives the decomposition into the two sectors
$x_1>x_2$ and $x_2>x_1$.
Eq.~(\ref{substitution}) transforms the triangles into squares.
This transformation is one-to-one for all points except
the origin. The origin is replaced by the line $x_1=0$ in the first sector
and by the line $x_2=0$ in the second sector.
Therefore the name ``blow-up''.
\\
\\
Step 3: The singular behaviour of the integral depends now only on the factor
\bq
  \prod\limits_{i=1}^{n}x_{i}^{a_{i}+\epsilon b_{i}}.
\eq
We Taylor expand in the integration variables and perform the trivial integrations
\bq
 \int\limits_{0}^{1} dx \; x^{a+b\eps}
 & = &
 \frac{1}{a+1+b\eps},
\eq
leading to the explicit poles in $1/\eps$.
\\
\\
Step 4: All remaining integrals are now by construction finite.
We can now expand all expressions in a Laurent series in $\eps$
and truncate to the desired order.
\\
\\
Step 5: It remains to compute the coefficients of the Laurent series.
These coefficients contain finite integrals, which can be evaluated numerically
by Monte Carlo integration.
We implemented\footnote{The program can be obtained from {\tt http://www.higgs.de/\~{}stefanw/software.html}}
the algorithm into a computer program, which computes numerically the coefficients 
of the Laurent series of any multi-loop integral \cite{Bogner:2007cr}.

\section{Hironaka's polyhedra game}
\label{sect:hironaka}

In step 2 of the algorithm we have an iteration.
It is important to show that this iteration terminates and does not lead to an infinite
loop.
\begin{figure*}[t]
\includegraphics[bb= 100 570 700 710,width=\textwidth]{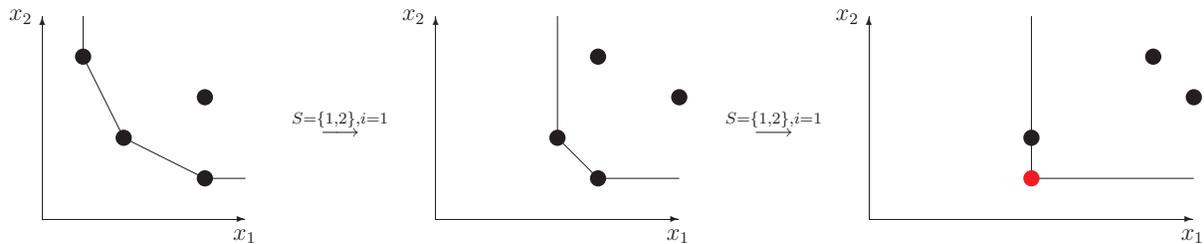}
\caption{\label{fig2} Illustration of Hironaka's polyhedra game.}
\end{figure*}
There are strategies for choosing the sub-sectors, which guarantee termination.
These strategies \cite{Hironaka:1964,Spivakovsky:1983,Encinas:2002,Hauser:2003,Zeillinger:2006} are closely related to Hironaka's polyhedra game.

Hironaka's polyhedra game is played by two players, A and B. They are
given a finite set $M$ of points $m=\left(m_{1},\,...,\,m_{n}\right)$
in $\mathbb{N}_{+}^{n}$, the first quadrant of $\mathbb{N}^{n}$.
We denote by $\Delta \subset\mathbb{R}_{+}^{n}$ the positive convex hull of the set $M$.
It is given by the convex hull of the set
\bq
\bigcup\limits_{m\in M}\left(m+\mathbb{R}_{+}^{n}\right).
\eq
The two players compete in the following game:
\begin{enumerate}
\item Player A chooses a non-empty subset $S\subseteq\left\{ 1,\,...,\, n\right\}$.
\item Player B chooses one element $i$ out of this subset $S$. 
\end{enumerate}
Then, according
to the choices of the players, the components of all $\left(m_{1},\,...,\,m_{n}\right)\in M$
are replaced by new points $\left(m_{1}^{\prime},\,...,\,m_{n}^{\prime}\right)$,
given by:
\bq
\label{update_polyhedron}
m_{j}^{\prime} & = & m_{j}, \;\;\; \textrm{if }j\neq i, \nonumber \\
m_{i}^{\prime} & = & \sum_{j\in S} m_{j}-c,
\eq
where for the moment we set $c=1$. 
This defines the set $M^\prime$.
One then sets $M=M^\prime$ and goes back to step 1.
Player A wins the game if, after a finite number of moves, 
the polyhedron $\Delta$ is of the form
\bq
\label{termination}
 \Delta & = & m+\mathbb{R}_{+}^{n},
\eq
i.e. generated by one point.
If this never occurs, player $B$ has won.
The challenge of the polyhedra game is to show that player $A$ always has
a winning strategy, no matter how player $B$ chooses his moves.
A simple illustration of Hironaka's polyhedra game in two dimensions is given in
fig.~\ref{fig2}. Player A always chooses $S=\{1,2\}$.
In ref.~\cite{Bogner:2007cr} we have shown that a winning strategy for 
Hironaka's polyhedra game
translates directly into a strategy for choosing the sub-sectors which
guarantees termination.

\section{Shuffle algebras}
\label{sect:shuffle}

Before we continue the discussion of loop integrals, it is useful to discuss first
shuffle algebras and generalisations thereof from an algebraic viewpoint.
Consider a set of letters $A$. The set $A$ is called the alphabet.
A word is an ordered sequence of letters:
\bq
 w & = & l_1 l_2 ... l_k.
\eq
The word of length zero is denoted by $e$.
Let $K$ be a field and consider the vector space of words over $K$.
A shuffle algebra ${\cal A}$ on the vector space of words is defined by
\bq
\left( l_1 l_2 ... l_k \right) \cdot 
 \left( l_{k+1} ... l_r \right) & = &
 \sum\limits_{\mbox{\tiny shuffles} \; \sigma} l_{\sigma(1)} l_{\sigma(2)} ... l_{\sigma(r)},
\eq
where the sum runs over all permutations $\sigma$, which preserve the relative order
of $1,2,...,k$ and of $k+1,...,r$.
The name ``shuffle algebra'' is related to the analogy of shuffling cards: If a deck of cards
is split into two parts and then shuffled, the relative order within the two individual parts
is conserved.
A shuffle algebra is also known under the name ``mould symmetral'' \cite{Ecalle}.
The empty word $e$ is the unit in this algebra:
\bq
 e \cdot w = w \cdot e = w.
\eq
A recursive definition of the shuffle product is given by
\bq
\label{def_recursive_shuffle}
\left( l_1 l_2 ... l_k \right) \cdot \left( l_{k+1} ... l_r \right) & = &
 l_1 \left[ \left( l_2 ... l_k \right) \cdot \left( l_{k+1} ... l_r \right) \right]
+
 l_{k+1} \left[ \left( l_1 l_2 ... l_k \right) \cdot \left( l_{k+2} ... l_r \right) \right].
\eq
It is well known fact that the shuffle algebra is actually a (non-cocommutative) Hopf algebra \cite{Reutenauer}.
In this context let us briefly review the definitions of a coalgebra, a bialgebra and a Hopf algebra,
which are closely related:
First note that the unit in an algebra can be viewed as a map from $K$ to $A$ and that the multiplication
can be viewed as a map from the tensor product $A \otimes A$ to $A$ (e.g. one takes two elements
from $A$, multiplies them and gets one element out). 

A coalgebra has instead of multiplication and unit the dual structures:
a comultiplication $\Delta$ and a counit $\bar{e}$.
The counit is a map from $A$ to $K$, whereas comultiplication is a map from $A$ to
$A \otimes A$.
Note that comultiplication and counit go in the reverse direction compared to multiplication
and unit.
We will always assume that the comultiplication is coassociative.
The general form of the coproduct is
\bq
\Delta(a) & = & \sum\limits_i a_i^{(1)} \otimes a_i^{(2)},
\eq
where $a_i^{(1)}$ denotes an element of $A$ appearing in the first slot of $A \otimes A$ and
$a_i^{(2)}$ correspondingly denotes an element of $A$ appearing in the second slot.
Sweedler's notation \cite{Sweedler} consists in dropping the dummy index $i$ and the summation symbol:
\bq
\Delta(a) & = & 
a^{(1)} \otimes a^{(2)}
\eq 
The sum is implicitly understood. This is similar to Einstein's summation convention, except
that the dummy summation index $i$ is also dropped. The superscripts ${}^{(1)}$ and ${}^{(2)}$ 
indicate that a sum is involved.

A bialgebra is an algebra and a coalgebra at the same time,
such that the two structures are compatible with each other.
Using Sweedler's notation,
the compatibility between the multiplication and comultiplication is express\-ed as
\bq
\label{bialg}
 \Delta\left( a \cdot b \right)
 & = &
\left( a^{(1)} \cdot b^{(1)} \right)
 \otimes \left( a^{(2)} \cdot b^{(2)} \right).
\eq

A Hopf algebra is a bialgebra with an additional map from $A$ to $A$, called the 
antipode ${\cal S}$, which fulfils
\bq
a^{(1)} \cdot {\cal S}\left( a^{(2)} \right)
=
{\cal S}\left(a^{(1)}\right) \cdot a^{(2)} 
= 0 & &\;\;\; \mbox{for} \; a \neq e.
\eq

With this background at hand we can now state the coproduct, the counit and the antipode for the
shuffle algebra:
The counit $\bar{e}$ is given by:
\bq
\bar{e}\left( e\right) = 1, \;\;\;
& &
\bar{e}\left( l_1 l_2 ... l_n\right) = 0.
\eq
The coproduct $\Delta$ is given by:
\bq
\Delta\left( l_1 l_2 ... l_k \right) 
& = & \sum\limits_{j=0}^k \left( l_{j+1} ... l_k \right) \otimes \left( l_1 ... l_j \right).
\eq
The antipode ${\cal S}$ is given by:
\bq
{\cal S}\left( l_1 l_2 ... l_k \right) & = & (-1)^k \; l_k l_{k-1} ... l_2 l_1.
\eq
The shuffle algebra is generated by the Lyndon words.
If one introduces a lexicographic ordering on the letters of the alphabet
$A$, a Lyndon word is defined by the property
\bq
w < v
\eq
for any sub-words $u$ and $v$ such that $w= u v$.

\begin{figure}
\includegraphics[bb= 70 640 535 720]{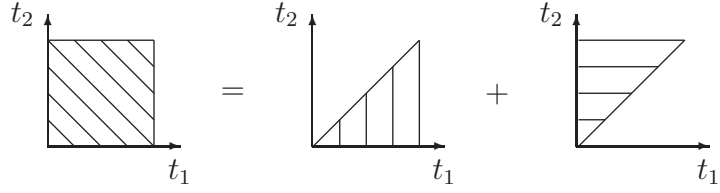}
\caption{\label{proof_shuffle} Sketch of the proof for the shuffle product of two iterated integrals.
The integral over the square is replaced by two
integrals over the upper and lower triangle.}
\end{figure}

An important example for a shuffle algebra are iterated integrals.
Let $[a, b]$ be a segment of the real line and $f_1$, $f_2$, ... functions
on this interval.
Let us define the following iterated integrals:
\bq
 I(f_1,f_2,...,f_k;a,b) 
 & = &
 \int\limits_a^b dt_1 f_1(t_1) \int\limits_a^{t_1} dt_2 f_2(t_2) 
 ...
 \int\limits_a^{t_{k-1}} dt_k f_k(t_k) 
\eq
For fixed $a$ and $b$ we have a shuffle algebra:
\bq
 I(f_1,f_2,...,f_k;a,b) \cdot I(f_{k+1},...,f_r; a,b) & = &
 \sum\limits_{\mbox{\tiny shuffles} \; \sigma} I(f_{\sigma(1)},f_{\sigma(2)},...,f_{\sigma(r)};a,b),
\eq
where the sum runs over all permutations $\sigma$, which preserve the relative order
of $1,2,...,k$ and of $k+1,...,r$.
The proof is sketched in fig.~\ref{proof_shuffle}.
The two outermost integrations are recursively replaced by integrations over the upper and lower triangle.

We now consider generalisations of shuffle algebras. Assume that for the set of letters we have an additional
operation
\bq
 (.,.) & : & A \otimes A \rightarrow A,
 \nonumber \\
       & &  l_1 \otimes l_2 \rightarrow (l_1, l_2),
\eq
which is commutative and associative.
Then we can define a new product of words recursively through
\bq
\label{def_recursive_quasi_shuffle}
\left( l_1 l_2 ... l_k \right) \ast \left( l_{k+1} ... l_r \right) & = &
 l_1 \left[ \left( l_2 ... l_k \right) \ast \left( l_{k+1} ... l_r \right) \right]
+
 l_{k+1} \left[ \left( l_1 l_2 ... l_k \right) \ast \left( l_{k+2} ... l_r \right) \right]
 \nonumber \\
 & &
+
(l_1,l_{k+1}) \left[ \left( l_2 ... l_k \right) \ast \left( l_{k+2} ... l_r \right) \right].
\eq
This product is a generalisation of the shuffle product and differs from the recursive
definition of the shuffle product in eq.~(\ref{def_recursive_shuffle}) through the extra term in the last line.
This modified product is known under the names quasi-shuffle product \cite{Hoffman},
mixable shuffle product \cite{Guo},
stuffle product \cite{Borwein} or
mould symmetrel \cite{Ecalle}.
Quasi-shuffle algebras are Hopf algebras.
Comultiplication and counit are defined as for the shuffle algebras.
The counit $\bar{e}$ is given by:
\bq
\bar{e}\left( e\right) = 1, \;\;\;
& &
\bar{e}\left( l_1 l_2 ... l_n\right) = 0.
\eq
The coproduct $\Delta$ is given by:
\bq
\Delta\left( l_1 l_2 ... l_k \right) 
& = & \sum\limits_{j=0}^k \left( l_{j+1} ... l_k \right) \otimes \left( l_1 ... l_j \right).
\eq
The antipode ${\cal S}$ is recursively defined through
\bq
{\cal S}\left( l_1 l_2 ... l_k \right) & = & 
 - l_1 l_2 ... l_k
 - \sum\limits_{j=1}^{k-1} {\cal S}\left( l_{j+1} ... l_k \right) \ast \left( l_1 ... l_j \right).
\eq
An example for a quasi-shuffle algebra are nested sums.
Let $n_a$ and $n_b$ be integers with $n_a<n_b$ and let $f_1$, $f_2$, ... be functions
defined on the integers.
We consider the following nested sums:
\bq
 S(f_1,f_2,...,f_k;n_a,n_b) 
 & = &
 \sum\limits_{i_1=n_a}^{n_b} f_1(i_1) \sum\limits_{i_2=n_a}^{i_1-1} f_2(i_2) 
 ...
 \sum\limits_{i_k=n_a}^{i_{k-1}-1} f_k(i_k)
\eq
For fixed $n_a$ and $n_b$ we have a quasi-shuffle algebra:
\bq
\label{quasi_shuffle_multiplication}
\lefteqn{
 S(f_1,f_2,...,f_k;n_a,n_b) \ast S(f_{k+1},...,f_r; n_a,n_b) 
= } & &
 \nonumber \\
 & &
   \sum\limits_{i_1=n_a}^{n_b} f_1(i_1) \; S(f_2,...,f_k;n_a,i_1-1) \ast S(f_{k+1},...,f_r; n_a,i_1-1)
 \nonumber \\
 & &
 +  \sum\limits_{j_1=n_a}^{n_b} f_k(j_1) \; S(f_1,f_2,...,f_k;n_a,j_1-1) \ast S(f_{k+2},...,f_r; n_a,j_1-1)
 \nonumber \\
 & &
 +  \sum\limits_{i=n_a}^{n_b} f_1(i) f_k(i) \; S(f_2,...,f_k;n_a,i-1) \ast S(f_{k+2},...,f_r; n_a,i-1)
\eq
\begin{figure}
\includegraphics[bb= 65 630 530 710]{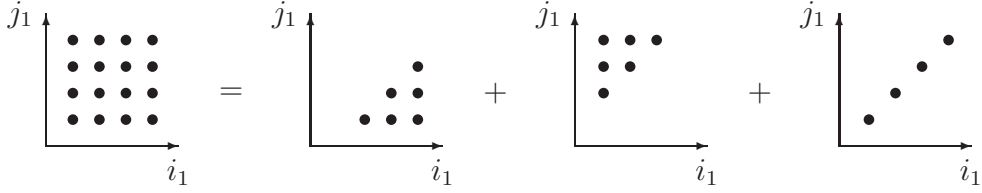}
\caption{\label{proof} Sketch of the proof for the quasi-shuffle product of nested sums. 
The sum over the square is replaced by
the sum over the three regions on the r.h.s.}
\end{figure}
Note that the product of two letters corresponds to the point-wise product of the two functions:
\bq
 ( f_i, f_j ) \; (n) & = & f_i(n) f_j(n).
\eq
The proof that nested sums obey the quasi-shuffle algebra is sketched in Fig. \ref{proof}.
The outermost sums of the nested sums on the l.h.s of (\ref{quasi_shuffle_multiplication}) are split into the three
regions indicated in Fig. \ref{proof}.

\section{Multiple polylogarithms}
\label{sect:polylog}

In the previous section we have seen that iterated integrals form a shuffle algebra, while
nested sums form a quasi-shuffle algebra.
In this context multiple polylogarithms form an interesting class of functions.
They have a representation as iterated integrals as well as nested sums.
Therefore multiple polylogarithms form a shuffle algebra as well as a quasi-shuffle algebra.
The two algebra structures are independent.
Let us start with the representation as nested sums.
The multiple polylogarithms are defined by \cite{Goncharov,Minh:2000,Cartier:2001,Racinet:2002}
\bq 
\label{multipolylog2}
 \mbox{Li}_{m_1,...,m_k}(x_1,...,x_k)
  & = & \sum\limits_{i_1>i_2>\ldots>i_k>0}
     \frac{x_1^{i_1}}{{i_1}^{m_1}}\ldots \frac{x_k^{i_k}}{{i_k}^{m_k}}.
\eq
The multiple polylogarithms are generalisations of
the classical polylogarithms 
$\mbox{Li}_n(x)$,
whose most prominent examples are
\bq
 \mbox{Li}_1(x) = \sum\limits_{i_1=1}^\infty \frac{x^{i_1}}{i_1} = -\ln(1-x),
 & &
 \mbox{Li}_2(x) = \sum\limits_{i_1=1}^\infty \frac{x^{i_1}}{i_1^2},
\eq 
as well as
Nielsen's generalised polylogarithms \cite{Nielsen}
\bq
S_{n,p}(x) & = & \mbox{Li}_{n+1,1,...,1}(x,\underbrace{1,...,1}_{p-1}),
\eq
and the harmonic polylogarithms \cite{Remiddi:1999ew,Gehrmann:2000zt}
\bq
\label{harmpolylog}
H_{m_1,...,m_k}(x) & = & \mbox{Li}_{m_1,...,m_k}(x,\underbrace{1,...,1}_{k-1}).
\eq

In addition, multiple polylogarithms have an integral representation. 
To discuss the integral representation it is convenient to 
introduce for $z_k \neq 0$
the following functions
\bq
\label{Gfuncdef}
G(z_1,...,z_k;y) & = &
 \int\limits_0^y \frac{dt_1}{t_1-z_1}
 \int\limits_0^{t_1} \frac{dt_2}{t_2-z_2} ...
 \int\limits_0^{t_{k-1}} \frac{dt_k}{t_k-z_k}.
\eq
In this definition 
one variable is redundant due to the following scaling relation:
\bq
G(z_1,...,z_k;y) & = & G(x z_1, ..., x z_k; x y)
\eq
If one further defines
\bq
g(z;y) & = & \frac{1}{y-z},
\eq
then one has
\bq
\label{derivative}
\frac{d}{dy} G(z_1,...,z_k;y) & = & g(z_1;y) G(z_2,...,z_k;y)
\eq
and
\bq
\label{Grecursive}
G(z_1,z_2,...,z_k;y) & = & \int\limits_0^y dt \; g(z_1;t) G(z_2,...,z_k;t).
\eq
One can slightly enlarge the set and define
$G(0,...,0;y)$ with $k$ zeros for $z_1$ to $z_k$ to be
\bq
\label{trailingzeros}
G(0,...,0;y) & = & \frac{1}{k!} \left( \ln y \right)^k.
\eq
This permits us to allow trailing zeros in the sequence
$(z_1,...,z_k)$ by defining the function $G$ with trailing zeros via (\ref{Grecursive}) 
and (\ref{trailingzeros}).
To relate the multiple polylogarithms to the functions $G$ it is convenient to introduce
the following short-hand notation:
\bq
\label{Gshorthand}
G_{m_1,...,m_k}(z_1,...,z_k;y)
 & = &
 G(\underbrace{0,...,0}_{m_1-1},z_1,...,z_{k-1},\underbrace{0...,0}_{m_k-1},z_k;y)
\eq
Here, all $z_j$ for $j=1,...,k$ are assumed to be non-zero.
One then finds
\bq
\label{Gintrepdef}
\mbox{Li}_{m_1,...,m_k}(x_1,...,x_k)
& = & (-1)^k 
 G_{m_1,...,m_k}\left( \frac{1}{x_1}, \frac{1}{x_1 x_2}, ..., \frac{1}{x_1...x_k};1 \right).
\eq
The inverse formula reads
\bq
G_{m_1,...,m_k}(z_1,...,z_k;y) & = & 
 (-1)^k \; \mbox{Li}_{m_1,...,m_k}\left(\frac{y}{z_1}, \frac{z_1}{z_2}, ..., \frac{z_{k-1}}{z_k}\right).
\eq
Eq. (\ref{Gintrepdef}) together with 
(\ref{Gshorthand}) and (\ref{Gfuncdef})
defines an integral representation for the multiple polylogarithms.

Up to now we treated multiple polylogarithms from an algebraic point of view.
Equally important are the analytical properties, which are needed for an efficient numerical 
evaluation.
As an example I first discuss the numerical evaluation of the dilogarithm \cite{'tHooft:1979xw}:
\bq
\mbox{Li}_{2}(x) & = & - \int\limits_{0}^{x} dt \frac{\ln(1-t)}{t}
 = \sum\limits_{n=1}^{\infty} \frac{x^{n}}{n^{2}}
\eq
The power series expansion can be evaluated numerically, provided $|x| < 1.$
Using the functional equations 
\bq
\mbox{Li}_2(x) & = & -\mbox{Li}_2\left(\frac{1}{x}\right) -\frac{\pi^2}{6} -\frac{1}{2} \left( \ln(-x) \right)^2,
 \nonumber \\
\mbox{Li}_2(x) & = & -\mbox{Li}_2(1-x) + \frac{\pi^2}{6} -\ln(x) \ln(1-x).
\eq
any argument of the dilogarithm can be mapped into the region
$|x| \le 1$ and
$-1 \leq \mbox{Re}(x) \leq 1/2$.
The numerical computation can be accelerated  by using an expansion in $[-\ln(1-x)]$ and the
Bernoulli numbers $B_i$:
\bq
\mbox{Li}_2(x) & = & \sum\limits_{i=0}^\infty \frac{B_i}{(i+1)!} \left( - \ln(1-x) \right)^{i+1}.
\eq
The generalisation to multiple polylogarithms proceeds along the same lines \cite{Vollinga:2004sn}:
Using the integral representation eq.~(\ref{Gfuncdef})
one transforms all arguments into a region, 
where one has a converging power series expansion.
In this region eq.~(\ref{multipolylog2}) may be used.
However it is advantageous to speed up the convergence of the power series expansion.
This is done as follows: 
The multiple polylogarithms satisfy the H\"older convolution \cite{Borwein}.
For $z_1 \neq 1$ and $z_w \neq 0$ this identity reads
\bq
\label{defhoelder}
\lefteqn{
G\left(z_1,...,z_w; 1 \right) 
 = } & & 
 \\
 & &
 \sum\limits_{j=0}^w \left(-1\right)^j 
  G\left(1-z_j, 1-z_{j-1},...,1-z_1; 1 - \frac{1}{p} \right)
  G\left( z_{j+1},..., z_w; \frac{1}{p} \right).
 \nonumber 
\eq
The H\"older convolution can be used to accelerate the 
convergence for the series
representation of the multiple polylogarithms.

\section{From Feynman integrals to multiple polylogarithms}
\label{sect:calc}

In sect.~\ref{sect:multi_loop} we saw that the Feynman parameter integrals 
depend on two graph polynomials ${\mathcal U}$ and ${\mathcal F}$, which are homogeneous functions of the 
Feynman parameters.
In this section we will discuss how multiple polylogarithms arise in the calculation of Feynman parameter
integrals.
We will discuss two approaches. In the first approach one uses a Mellin-Barnes transformation and sums
up residues. This leads to the sum representation of multiple polylogarithms.
In the second approach one first derives a differential equation for the Feynman parameter integral, which
is then solved by an ansatz in terms of the iterated integral representation of multiple polylogarithms.

Let us start with the first approach. Assume for the moment that the two graph polynomials 
${\mathcal U}$ and ${\mathcal F}$ are absent from the Feynman parameter integral.
In this case we have
\bq
\label{multi_beta_fct}
 \int\limits_{0}^{1} \left( \prod\limits_{j=1}^{n}\,dx_j\,x_j^{\nu_j-1} \right)
 \delta(1-\sum_{i=1}^n x_i)
 & = & 
 \frac{\prod\limits_{j=1}^{n}\Gamma(\nu_j)}{\Gamma(\nu_1+...+\nu_n)}.
\eq
With the help of 
the Mellin-Barnes transformation we now reduce the general case to eq.~(\ref{multi_beta_fct}).
The Mellin-Barnes transformation reads
\bq
\label{multi_mellin_barnes}
\lefteqn{
\left(A_1 + A_2 + ... + A_n \right)^{-c} 
 = 
 \frac{1}{\Gamma(c)} \frac{1}{\left(2\pi i\right)^{n-1}} 
 \int\limits_{-i\infty}^{i\infty} d\sigma_1 ... \int\limits_{-i\infty}^{i\infty} d\sigma_{n-1}
 } & & \\
 & & 
 \times 
 \Gamma(-\sigma_1) ... \Gamma(-\sigma_{n-1}) \Gamma(\sigma_1+...+\sigma_{n-1}+c)
 \; 
 A_1^{\sigma_1} ...  A_{n-1}^{\sigma_{n-1}} A_n^{-\sigma_1-...-\sigma_{n-1}-c}.
 \nonumber 
\eq
Each contour is such that the poles of $\Gamma(-\sigma)$ are to the right and the poles
of $\Gamma(\sigma+c)$ are to the left.
This transformation can be used to convert the sum of monomials of the polynomials ${\mathcal U}$ and ${\mathcal F}$ into
a product, such that all Feynman parameter integrals are of the form of eq.~(\ref{multi_beta_fct}).
As this transformation converts sums into products it is 
the ``inverse'' of Feynman parametrisation.

With the help of eq.~(\ref{multi_beta_fct}) and eq.~(\ref{multi_mellin_barnes})
we may exchange the Feynman parameter integrals against multiple contour integrals.
A single contour integral is of the form
\bq
\label{MellinBarnesInt}
I
 & = & 
\frac{1}{2\pi i} \int\limits_{\gamma-i\infty}^{\gamma+i\infty}
 d\sigma \; 
 \frac{\Gamma(\sigma+a_1) ... \Gamma(\sigma+a_m)}
      {\Gamma(\sigma+c_2) ... \Gamma(\sigma+c_p)}
 \frac{\Gamma(-\sigma+b_1) ... \Gamma(-\sigma+b_n)}
      {\Gamma(-\sigma+d_1) ... \Gamma(-\sigma+d_q)} 
 \; x^{-\sigma}.
\eq
If $\;\mbox{max}\left( \mbox{Re}(-a_1), ..., \mbox{Re}(-a_m) \right) < \mbox{min}\left( \mbox{Re}(b_1), ..., \mbox{Re}(b_n) \right)$ the contour can be chosen
as a straight line parallel to the imaginary axis with
\bq
\mbox{max}\left( \mbox{Re}(-a_1), ..., \mbox{Re}(-a_m) \right) 
 \;\;\; < \;\;\; \mbox{Re} \; \gamma \;\;\; < \;\;\;
\mbox{min}\left( \mbox{Re}(b_1), ..., \mbox{Re}(b_n) \right),
\eq
otherwise the contour is indented, such that the residues of
$\Gamma(\sigma+a_1)$, ..., $\Gamma(\sigma+a_m)$ are to the right of the contour,
whereas the residues of 
$\Gamma(-\sigma+b_1)$,  ..., $\Gamma(-\sigma+b_n)$ are to the left of the contour.
The integral eq. (\ref{MellinBarnesInt}) is most conveniently evaluated with 
the help of the residuum theorem by closing the contour to the left or to the right.
To sum up all residues which lie inside the contour
it is useful to know the residues of the Gamma function:
\bq
\mbox{res} \; \left( \Gamma(\sigma+a), \sigma=-a-n \right) = \frac{(-1)^n}{n!}, 
 & &
\mbox{res} \; \left( \Gamma(-\sigma+a), \sigma=a+n \right) = -\frac{(-1)^n}{n!}. 
\eq
In general there are multiple contour integrals, and as a consequence one obtains multiple sums.
Having collected all residues, one then expands the Gamma-functions:
\bq
\label{expansiongamma}
\lefteqn{
\Gamma(n+\eps)  = 
} & & \\
 & & \Gamma(1+\eps) \Gamma(n)
 \left[
        1 + \eps Z_1(n-1) + \eps^2 Z_{11}(n-1)
          + \eps^3 Z_{111}(n-1) + ... + \eps^{n-1} Z_{11...1}(n-1)
 \right],
 \nonumber
\eq
where $Z_{m_1,...,m_k}(n)$ are Euler-Zagier sums
defined by
\bq
 Z_{m_1,...,m_k}(n) & = &
  \sum\limits_{n \ge i_1>i_2>\ldots>i_k>0}
     \frac{1}{{i_1}^{m_1}}\ldots \frac{1}{{i_k}^{m_k}}.
\eq
This motivates the following definition of a special form of nested sums, called 
$Z$-sums \cite{Moch:2001zr,Weinzierl:2002hv,Weinzierl:2004bn,Moch:2005uc}:
\bq 
\label{definition}
  Z(n;m_1,...,m_k;x_1,...,x_k) & = & \sum\limits_{n\ge i_1>i_2>\ldots>i_k>0}
     \frac{x_1^{i_1}}{{i_1}^{m_1}}\ldots \frac{x_k^{i_k}}{{i_k}^{m_k}}.
\eq
$k$ is called the depth of the $Z$-sum and $w=m_1+...+m_k$ is called the weight.
If the sums go to infinity ($n=\infty$) the $Z$-sums are multiple polylogarithms:
\bq
\label{multipolylog}
Z(\infty;m_1,...,m_k;x_1,...,x_k) & = & \mbox{Li}_{m_1,...,m_k}(x_1,...,x_k).
\eq
For $x_1=...=x_k=1$ the definition reduces to the Euler-Zagier sums \cite{Euler,Zagier,Vermaseren:1998uu,Blumlein:1998if,Blumlein:2003gb}:
\bq
Z(n;m_1,...,m_k;1,...,1) & = & Z_{m_1,...,m_k}(n).
\eq
For $n=\infty$ and $x_1=...=x_k=1$ the sum is a multiple $\zeta$-value \cite{Borwein,Blumlein:2009}:
\bq
Z(\infty;m_1,...,m_k;1,...,1) & = & \zeta_{m_1,...,m_k}.
\eq
The usefulness of the $Z$-sums lies in the fact, that they interpolate between
multiple polylogarithms and Euler-Zagier sums.
The $Z$-sums form a quasi-shuffle algebra.
In this approach multiple polylogarithms appear through eq.~(\ref{multipolylog}).

An alternative approach to the computation of Feynman parameter integrals is based on
differential equations \cite{Kotikov:1990kg,Kotikov:1991pm,Remiddi:1997ny,Gehrmann:1999as,Gehrmann:2000zt,Gehrmann:2001ck}.
To evaluate these integrals within this approach 
one first finds for each 
master integral a differential
equation, which this master integral has to satisfy.
The derivative is taken with respect to an external scale, or a
ratio of two scales.
An example for a one-loop four-point function is given by
\bq
\lefteqn{
\frac{\partial}{\partial s_{123}}
\begin{picture}(140,40)(-15,45)
\Vertex(50,20){2}
\Vertex(50,80){2}
\Vertex(20,50){2}
\Vertex(80,50){2}
\Line(0,50)(20,50)
\Line(20,50)(50,80)
\Line(50,20)(20,50)
\Line(50,80)(80,50)
\Line(80,50)(50,20)
\Line(50,80)(70,80)
\Line(80,50)(100,50)
\Line(50,20)(70,20)
\Text(75,80)[l]{\tiny $p_1$}
\Text(105,50)[l]{\tiny $p_2$}
\Text(75,20)[l]{\tiny $p_3$}
\end{picture}
= 
\frac{D-4}{2(s_{12}+s_{23}-s_{123})}
\begin{picture}(100,40)(-15,45)
\Vertex(50,20){2}
\Vertex(50,80){2}
\Vertex(20,50){2}
\Vertex(80,50){2}
\Line(0,50)(20,50)
\Line(20,50)(50,80)
\Line(50,20)(20,50)
\Line(50,80)(80,50)
\Line(80,50)(50,20)
\Line(50,80)(70,80)
\Line(80,50)(100,50)
\Line(50,20)(70,20)
\Text(75,80)[l]{\tiny $p_1$}
\Text(105,50)[l]{\tiny $p_2$}
\Text(75,20)[l]{\tiny $p_3$}
\end{picture}
} & &
\nonumber \\
& & \nonumber \\
 & &
 + \frac{2(D-3)}{(s_{123}-s_{12})(s_{123}-s_{12}-s_{23})}
\left[ 
\frac{1}{s_{123}}
\begin{picture}(110,40)(-5,45)
\Vertex(30,50){2}
\Vertex(70,50){2}
\Line(10,50)(30,50)
\CArc(50,50)(20,0,360)
\Line(70,50)(90,50)
\Text(80,55)[lb]{\tiny $p_{123}$}
\end{picture}
-
\frac{1}{s_{12}}
\begin{picture}(110,40)(-5,45)
\Vertex(30,50){2}
\Vertex(70,50){2}
\Line(10,50)(30,50)
\CArc(50,50)(20,0,360)
\Line(70,50)(90,50)
\Text(80,55)[lb]{\tiny $p_{12}$}
\end{picture}
\right]
\nonumber \\
 & &
 + \frac{2(D-3)}{(s_{123}-s_{23})(s_{123}-s_{12}-s_{23})}
\left[ 
\frac{1}{s_{123}}
\begin{picture}(110,40)(-5,45)
\Vertex(30,50){2}
\Vertex(70,50){2}
\Line(10,50)(30,50)
\CArc(50,50)(20,0,360)
\Line(70,50)(90,50)
\Text(80,55)[lb]{\tiny $p_{123}$}
\end{picture}
-
\frac{1}{s_{23}}
\begin{picture}(110,40)(-5,45)
\Vertex(30,50){2}
\Vertex(70,50){2}
\Line(10,50)(30,50)
\CArc(50,50)(20,0,360)
\Line(70,50)(90,50)
\Text(80,55)[lb]{\tiny $p_{23}$}
\end{picture}
\right].
\nonumber
\eq
The two-point functions on the r.h.s are simpler and can be considered to be known.
This equation is solved iteratively by an ansatz 
for the solution as a Laurent expression in $\eps$.
Each term in this Laurent series is a sum of terms, consisting of
basis functions times
some unknown (and to be determined) coefficients.
This ansatz is inserted into the differential equation and the unknown 
coefficients
are determined order by order from the differential equation.
The basis functions are taken as a subset of multiple polylogarithms.
In this approach the iterated integral representation of multiple polylogarithms is the most
convenient form. This is immediately clear from the
simple formula for the derivative as in eq.~(\ref{derivative}).

\section{Conclusions}
\label{sect:conclusions}

In this talk we reported on mathematical properties of Feynman integrals.
We first showed that under rather weak assumptions all the coefficients 
of the Laurent expansion of a multi-loop integral
are periods.
In the second part we focused on multiple polylogarithms and how they appear in the calculation
of Feynman integrals.

\bibliography{/home/stefanw/notes/biblio}

\begin{thebibliography}{10}

\bibitem{'tHooft:1972fi}
G.~'t~Hooft and M.~J.~G. Veltman,
\newblock Nucl. Phys. {\bf B44}, 189 (1972).

\bibitem{Bollini:1972ui}
C.~G. Bollini and J.~J. Giambiagi,
\newblock Nuovo Cim. {\bf B12}, 20 (1972).

\bibitem{Cicuta:1972jf}
G.~M. Cicuta and E.~Montaldi,
\newblock Nuovo Cim. Lett. {\bf 4}, 329 (1972).

\bibitem{Moch:2004pa}
S.~Moch, J.~A.~M. Vermaseren, and A.~Vogt,
\newblock Nucl. Phys. {\bf B688}, 101 (2004), hep-ph/0403192.

\bibitem{Vogt:2004mw}
A.~Vogt, S.~Moch, and J.~A.~M. Vermaseren,
\newblock Nucl. Phys. {\bf B691}, 129 (2004), hep-ph/0404111.

\bibitem{Garland:2001tf}
L.~W. Garland, T.~Gehrmann, E.~W.~N. Glover, A.~Koukoutsakis, and E.~Remiddi,
\newblock Nucl. Phys. {\bf B627}, 107 (2002), hep-ph/0112081.

\bibitem{Garland:2002ak}
L.~W. Garland, T.~Gehrmann, E.~W.~N. Glover, A.~Koukoutsakis, and E.~Remiddi,
\newblock Nucl. Phys. {\bf B642}, 227 (2002), hep-ph/0206067.

\bibitem{Moch:2002hm}
S.~Moch, P.~Uwer, and S.~Weinzierl,
\newblock Phys. Rev. {\bf D66}, 114001 (2002), hep-ph/0207043.

\bibitem{GehrmannDeRidder:2007bj}
A.~Gehrmann-De~Ridder, T.~Gehrmann, E.~W.~N. Glover, and G.~Heinrich,
\newblock Phys. Rev. Lett. {\bf 99}, 132002 (2007), 0707.1285.

\bibitem{GehrmannDeRidder:2007hr}
A.~Gehrmann-De~Ridder, T.~Gehrmann, E.~W.~N. Glover, and G.~Heinrich,
\newblock JHEP {\bf 12}, 094 (2007), 0711.4711.

\bibitem{GehrmannDeRidder:2008ug}
A.~Gehrmann-De~Ridder, T.~Gehrmann, E.~W.~N. Glover, and G.~Heinrich,
\newblock Phys. Rev. Lett. {\bf 100}, 172001 (2008), 0802.0813.

\bibitem{GehrmannDeRidder:2009dp}
A.~Gehrmann-De~Ridder, T.~Gehrmann, E.~W.~N. Glover, and G.~Heinrich,
\newblock JHEP {\bf 05}, 106 (2009), 0903.4658.

\bibitem{Weinzierl:2008iv}
S.~Weinzierl,
\newblock Phys. Rev. Lett. {\bf 101}, 162001 (2008), 0807.3241.

\bibitem{Weinzierl:2009ms}
S.~Weinzierl,
\newblock JHEP {\bf 06}, 041 (2009), 0904.1077.

\bibitem{Weinzierl:2009yz}
S.~Weinzierl,
\newblock Phys. Rev. {\bf D80}, 094018 (2009), 0909.5056.

\bibitem{Bloch:2005}
S.~Bloch, H.~Esnault, and D.~Kreimer,
\newblock Comm. Math. Phys. {\bf 267}, 181 (2006), math.AG/0510011.

\bibitem{Bloch:2008jk}
S.~Bloch and D.~Kreimer,
\newblock Commun. Num. Theor. Phys. {\bf 2}, 637 (2008), 0804.4399.

\bibitem{Bloch:2008}
S.~Bloch,
\newblock (2008), arXiv:0810.1313 [math.AG].

\bibitem{Brown:2008}
F.~Brown,
\newblock Commun. Math. Phys. {\bf 287}, 925 (2008), arXiv:0804.1660 [math.AG].

\bibitem{Brown:2009a}
F.~Brown,
\newblock (2009), arXiv:0910.0114 [math.AG].

\bibitem{Brown:2009b}
F.~Brown and K.~Yeats,
\newblock (2009), arXiv:0910.5429 [math-ph].

\bibitem{Schnetz:2008mp}
O.~Schnetz,
\newblock (2008), arXiv:0801.2856 [hep-th].

\bibitem{Schnetz:2009}
O.~Schnetz,
\newblock (2009), arXiv:0909.0905 [math.CO].

\bibitem{Aluffi:2008sy}
P.~Aluffi and M.~Marcolli,
\newblock (2008), arXiv:0807.1690 [hep-th].

\bibitem{Aluffi:2008rw}
P.~Aluffi and M.~Marcolli,
\newblock (2008), arXiv:0811.2514 [hep-th].

\bibitem{Aluffi:2009b}
P.~Aluffi and M.~Marcolli,
\newblock (2009), arXiv:0901.2107 [math.AG].

\bibitem{Aluffi:2009a}
P.~Aluffi and M.~Marcolli,
\newblock (2009), arXiv:0907.3225 [math-ph].

\bibitem{Bergbauer:2009yu}
C.~Bergbauer, R.~Brunetti, and D.~Kreimer,
\newblock (2009), arXiv:0908.0633 [hep-th].

\bibitem{Laporta:2002pg}
S.~Laporta,
\newblock Phys. Lett. {\bf B549}, 115 (2002), hep-ph/0210336.

\bibitem{Laporta:2004rb}
S.~Laporta and E.~Remiddi,
\newblock Nucl. Phys. {\bf B704}, 349 (2005), hep-ph/0406160.

\bibitem{Laporta:2008sx}
S.~Laporta,
\newblock Int. J. Mod. Phys. {\bf A23}, 5007 (2008), 0803.1007.

\bibitem{Bailey:2008ib}
D.~H. Bailey, J.~M. Borwein, D.~Broadhurst, and M.~L. Glasser,
\newblock (2008), arXiv:0801.0891 [hep-th].

\bibitem{Bierenbaum:2003ud}
I.~Bierenbaum and S.~Weinzierl,
\newblock Eur. Phys. J. {\bf C32}, 67 (2003), hep-ph/0308311.

\bibitem{Tarasov:1996br}
O.~V. Tarasov,
\newblock Phys. Rev. {\bf D54}, 6479 (1996), hep-th/9606018.

\bibitem{Tarasov:1997kx}
O.~V. Tarasov,
\newblock Nucl. Phys. {\bf B502}, 455 (1997), hep-ph/9703319.

\bibitem{Kontsevich:2001}
M.~Kontsevich and D.~Zagier,
\newblock in: B. Engquis and W. Schmid, editors, Mathematics unlimited - 2001
  and beyond , 771 (2001).

\bibitem{Yoshinaga:2008}
M.~Yoshinaga,
\newblock (2008), arXiv:0805.0349 [math.AG].

\bibitem{Friedrich:2005}
B.~Friedrich,
\newblock (2005), arXiv:math.AG/0506113.

\bibitem{Belkale:2003}
P.~Belkale and P.~Brosnan,
\newblock Int. Math. Res. Not. , 2655 (2003).

\bibitem{Bogner:2007mn}
C.~Bogner and S.~Weinzierl,
\newblock J. Math. Phys. {\bf 50}, 042302 (2009), 0711.4863.

\bibitem{Hepp:1966eg}
K.~Hepp,
\newblock Commun. Math. Phys. {\bf 2}, 301 (1966).

\bibitem{Roth:1996pd}
M.~Roth and A.~Denner,
\newblock Nucl. Phys. {\bf B479}, 495 (1996), hep-ph/9605420.

\bibitem{Binoth:2000ps}
T.~Binoth and G.~Heinrich,
\newblock Nucl. Phys. {\bf B585}, 741 (2000), hep-ph/0004013.

\bibitem{Binoth:2003ak}
T.~Binoth and G.~Heinrich,
\newblock Nucl. Phys. {\bf B680}, 375 (2004), hep-ph/0305234.

\bibitem{Bogner:2007cr}
C.~Bogner and S.~Weinzierl,
\newblock Comput. Phys. Commun. {\bf 178}, 596 (2008), 0709.4092.

\bibitem{Smirnov:2008py}
A.~V. Smirnov and M.~N. Tentyukov,
\newblock Comput. Phys. Commun. {\bf 180}, 735 (2009), 0807.4129.

\bibitem{Smirnov:2008aw}
A.~V. Smirnov and V.~A. Smirnov,
\newblock (2008), arXiv:0812.4700.

\bibitem{Hironaka:1964}
H.~Hironaka,
\newblock Ann. Math. {\bf 79}, 109 (1964).

\bibitem{Spivakovsky:1983}
M.~Spivakovsky,
\newblock Progr. Math. {\bf 36}, 419 (1983).

\bibitem{Encinas:2002}
S.~Encinas and H.~Hauser,
\newblock Comment. Math. Helv. {\bf 77}, 821 (2002).

\bibitem{Hauser:2003}
H.~Hauser,
\newblock Bull. Amer. Math. Soc. {\bf 40}, 323 (2003).

\bibitem{Zeillinger:2006}
D.~Zeillinger,
\newblock Enseign. Math. {\bf 52}, 143 (2006).

\bibitem{Ecalle}
J.~Ecalle,
\newblock (2002), (available at
  http://www.math.u-psud.fr/~biblio/ppo/2002/ppo2002-23.html ).

\bibitem{Reutenauer}
C.~Reutenauer,
\newblock {\em Free Lie Algebras} (Clarendon Press, Oxford, 1993).

\bibitem{Sweedler}
M.~Sweedler,
\newblock {\em Hopf Algebras} (Benjamin, New York, 1969).

\bibitem{Hoffman}
M.~E. Hoffman,
\newblock J. Algebraic Combin. {\bf 11}, 49 (2000), math.QA/9907173.

\bibitem{Guo}
L.~Guo and W.~Keigher,
\newblock Adv. in Math. {\bf 150}, 117 (2000), math.RA/0407155.

\bibitem{Borwein}
J.~M. Borwein, D.~M. Bradley, D.~J. Broadhurst, and P.~Lisonek,
\newblock Trans. Amer. Math. Soc. {\bf 353:3}, 907 (2001), math.CA/9910045.

\bibitem{Goncharov}
A.~B. Goncharov,
\newblock Math. Res. Lett. {\bf 5}, 497 (1998), (available at
  http://www.math.uiuc.edu/K-theory/0297).

\bibitem{Minh:2000}
H.~M. Minh, M.~Petitot, and J.~van~der Hoeven,
\newblock Discrete Math. {\bf 225:1-3}, 217 (2000).

\bibitem{Cartier:2001}
P.~Cartier,
\newblock S{\'e}minaire Bourbaki , 885 (2001).

\bibitem{Racinet:2002}
G.~Racinet,
\newblock Publ. Math. Inst. Hautes Études Sci. {\bf 95}, 185 (2002),
  math.QA/0202142.

\bibitem{Nielsen}
N.~Nielsen,
\newblock Nova Acta Leopoldina (Halle) {\bf 90}, 123 (1909).

\bibitem{Remiddi:1999ew}
E.~Remiddi and J.~A.~M. Vermaseren,
\newblock Int. J. Mod. Phys. {\bf A15}, 725 (2000), hep-ph/9905237.

\bibitem{Gehrmann:2000zt}
T.~Gehrmann and E.~Remiddi,
\newblock Nucl. Phys. {\bf B601}, 248 (2001), hep-ph/0008287.

\bibitem{'tHooft:1979xw}
G.~'t~Hooft and M.~J.~G. Veltman,
\newblock Nucl. Phys. {\bf B153}, 365 (1979).

\bibitem{Vollinga:2004sn}
J.~Vollinga and S.~Weinzierl,
\newblock Comput. Phys. Commun. {\bf 167}, 177 (2005), hep-ph/0410259.

\bibitem{Moch:2001zr}
S.~Moch, P.~Uwer, and S.~Weinzierl,
\newblock J. Math. Phys. {\bf 43}, 3363 (2002), hep-ph/0110083.

\bibitem{Weinzierl:2002hv}
S.~Weinzierl,
\newblock Comput. Phys. Commun. {\bf 145}, 357 (2002), math-ph/0201011.

\bibitem{Weinzierl:2004bn}
S.~Weinzierl,
\newblock J. Math. Phys. {\bf 45}, 2656 (2004), hep-ph/0402131.

\bibitem{Moch:2005uc}
S.~Moch and P.~Uwer,
\newblock Comput. Phys. Commun. {\bf 174}, 759 (2006), math-ph/0508008.

\bibitem{Euler}
L.~Euler,
\newblock Novi Comm. Acad. Sci. Petropol. {\bf 20}, 140 (1775).

\bibitem{Zagier}
D.~Zagier,
\newblock First European Congress of Mathematics, Vol. II, Birkhauser, Boston ,
  497 (1994).

\bibitem{Vermaseren:1998uu}
J.~A.~M. Vermaseren,
\newblock Int. J. Mod. Phys. {\bf A14}, 2037 (1999), hep-ph/9806280.

\bibitem{Blumlein:1998if}
J.~Bl{\"u}mlein and S.~Kurth,
\newblock Phys. Rev. {\bf D60}, 014018 (1999), hep-ph/9810241.

\bibitem{Blumlein:2003gb}
J.~Bl{\"u}mlein,
\newblock Comput. Phys. Commun. {\bf 159}, 19 (2004), hep-ph/0311046.

\bibitem{Blumlein:2009}
J.~Bl{\"u}mlein, D.~J. Broadhurst, and J.~A.~M. Vermaseren,
\newblock (2009), arXiv:0907.2557 [math-ph].

\bibitem{Kotikov:1990kg}
A.~V. Kotikov,
\newblock Phys. Lett. {\bf B254}, 158 (1991).

\bibitem{Kotikov:1991pm}
A.~V. Kotikov,
\newblock Phys. Lett. {\bf B267}, 123 (1991).

\bibitem{Remiddi:1997ny}
E.~Remiddi,
\newblock Nuovo Cim. {\bf A110}, 1435 (1997), hep-th/9711188.

\bibitem{Gehrmann:1999as}
T.~Gehrmann and E.~Remiddi,
\newblock Nucl. Phys. {\bf B580}, 485 (2000), hep-ph/9912329.

\bibitem{Gehrmann:2001ck}
T.~Gehrmann and E.~Remiddi,
\newblock Nucl. Phys. {\bf B601}, 287 (2001), hep-ph/0101124.

\end{thebibliography}
\bibliographystyle{/home/stefanw/latex-style/h-physrev3}

\end{document}